\documentclass[11pt]{article}
\begin{document}
\begin{flushright}
SJSU/TP-02-23\\
April 2002\end{flushright}
\vspace{1.7in}
\begin{center}\Large{\bf Do macroscopic properties dictate microscopic
             probabilities?}\\
\vspace{1cm}
\normalsize\ J. Finkelstein\footnote[1]{
        Participating Guest, Lawrence Berkeley National Laboratory\\
        \hspace*{\parindent}\hspace*{1em}
        e-mail: JLFINKELSTEIN@lbl.gov}\\
        Department of Physics\\
        San Jos\'{e} State University\\San Jos\'{e}, CA 95192, U.S.A
\end{center}
\begin{abstract}
Aharonov and Reznik have recently argued that
the form of the probabilistic predictions of
quantum theory can be seen to follow from properties of macroscopic systems.
An error in their argument is identified.
\end{abstract}
\newpage
Aharonov and Reznik (ref. \cite{AR}, hereafter AR) have recently suggested
that the form of the probabilistic predictions of
quantum theory can be seen to follow from properties of macroscopic systems.  
This suggestion is made in the
context of acceptance of the non-probabilistic parts of standard
quantum theory (in particular the requirement that the result of
a measurement of an observable must be one of the eigenvalues  of
the operator corresponding to that observable), as well as the 
requirement that quantum predictions will indeed be probabilistic;
the result that AR wish to derive is the specific form those probabilistic
predictions take.  The purpose of this note is not to question the 
standard quantum predictions which AR wish to derive from a consideration
of macroscopic properties - indeed, those standard predictions are also
required by Gleason's theorem \cite{G} and of course are supported by
an enormous amount of experimental evidence - but rather to examine
whether these predictions can indeed be said to follow from the 
considerations which AR present.

AR consider a collection of $N$ identically-prepared spin-$\frac{1}{2}$
particles, which are in the product state
\begin{equation}
|\Psi \rangle = |\psi \rangle _{1} \otimes |\psi \rangle _{2}\otimes \ldots
                \otimes |\psi \rangle _{N},
\end{equation}
where $ |\psi \rangle _{r}$ is the state of the $r$th particle which is
given, for any $r$, by
\begin{equation}
|\psi \rangle _{r} = c_{+}|+\rangle _{r} + c_{-}|-\rangle _{r},
\end{equation}
and where $|+\rangle _r$ and $|-\rangle _r$ are the eigenstates of the 
operator $(\sigma _x)_r $. AR define $M_x $ to be the operator for the
average value of the $x$-component of spin, that is,
\begin{equation}
M_{x} = \left( \frac{1}{N}\right) \sum_{r=1}^{N} (\sigma _{x})_{r}.
\end{equation} 
Now let $f_+$ denote the probability of obtaining, in a measurement
of $\sigma_x$ on a single particle, the outcome $+1$; the result which
AR wish to prove is that $f_+$ is equal to $|c_{+}|^2$.

If $N$ is large, then if one were to measure $M_x$ by measuring 
$\sigma_x$ for each particle, the law of large numbers would require
that the outcome would almost certainly be close to 
$f_{+}(+1)+(1-f_{+})(-1)$. Now let $\bar{\sigma}_x$ denote
$|c_{+}|^{2}(+1)+|c_{-}|^{2}(-1)$ (since $|c_{-}|^{2} = 1 - |c_{+}|^2$,
this means $\bar{\sigma}_x = 2 |c_{+}|^{2} - 1$).
If AR can show that, in a measurement
of $M_x$ on the state $|\Psi \rangle $ the outcome would almost
certainly be close to $\bar{\sigma}_x$, then they would have the result
which they want.

AR begin by reminding us (see also refs.\ \cite{Hartle}, \cite{FGG} )
that if we write
\begin{equation}
M_{x}|\Psi \rangle = \bar{\sigma}_x |\Psi \rangle + |\Delta \rangle,
\end{equation}
then the norm of $|\Delta \rangle$ vanishes in the limit 
$N \rightarrow \infty $.
This fact might be interpreted to mean that, for $N = \infty $,
$|\Psi \rangle $ is an eigenstate of $M_x$ with eigenvalue $\bar{\sigma}_x$,
which (according to the non-probabilistic part of standard quantum
mechanics) would imply that a measurement of $M_x$ on the state
$|\Psi \rangle $ would surely have the outcome $\bar{\sigma}_x$.
However (as AR carefully point out), $N$ is finite for any actual
system, and so, in order to obtain the desired result for $N$ large 
but finite, AR introduce a ``stability assumption''; they write
\begin{quote}
One way to proceed is to make an additional assumption which seems
natural for macroscopically large samples: {\em the results of physical
experiments are stable against small perturbations.}
\end{quote}
Having made this assumption,  they then argue
for their result in two ways. The first of these is based on the 
assertion that a small change in $|\Psi \rangle $ would make it an
{\em exact} eigenstate of $M_x$.  In the second way, they consider  
a model for the interaction which results in a measurement of $M_x$,
and apply their assumption to the post-measurement state of the
$N$-particle system entangled with the measurement device.
 
In this note
I point out that the first of these ways is in fact in error, and
also note that, given their ``stability assumption'', it is not
necessary to consider the post-measurement state at all.  
The first argument given by AR begins with the following passage:
\begin{quote}
\ldots for finite large $N$, the operator $M_x$ fails to be a precise 
eigenoperator of $|\Psi \rangle $\ldots However, by a small modification
of the state to $|\Psi \rangle +|\delta \Psi \rangle $
with magnitude $||\: |\delta \Psi \rangle \: || = O(1/N)$,
the perturbed state does become an exact eigenstate of $M_x$.
\end{quote}
However, some time ago Squires \cite{S} showed that, 
as $N$ becomes large, $|\Psi \rangle $ becomes
{\em orthogonal} to any eigenstate of $M_x$. So, if we let $|M_{x}=m\rangle$
denote any (normalized) eigenstate of $M_x$, and write
\begin{equation}
|M_{x}=m\rangle = |\Psi \rangle +|\delta \Psi \rangle,
\end{equation}
we have
\begin{equation}
||\: |\delta \Psi \rangle \: ||^{2} = || \: |\Psi \rangle \: ||^{2}
   + || \: |M_{x}=m\rangle \: ||^{2} - 2{\rm Re}\langle M_{x}=m|\Psi\rangle,
\end{equation}
and since \cite{S} $\lim_{N \rightarrow \infty}
||\: \langle M_{x}=m|\Psi\rangle \: || = 0$,
this gives $\lim_{N \rightarrow \infty}|| \:\langle 
\delta \Psi \rangle \: ||= 2$,
contradicting the assertion of AR that 
$|| \:\langle \delta \Psi \rangle \: ||$ is $O(1/N)$. 

This result of Squires can be understood as follows: let $S$ be a
subset of the integers from $1$ to $N$, and define
\begin{equation}
|S\rangle = \left[ \bigotimes_{r\in S}|+\rangle _{r} \right]
             \left[ \bigotimes_{r\not \in S}|-\rangle _{r} \right]
\end{equation}
that is, $|S\rangle $ is the product N-particle state in which the $r$th
particle is in state $|+\rangle$ if $r$ is included in the set $S$,
and is in state $|-\rangle$ if $r$ is not included in the set $S$.
Let $n(S)$ be the number of elements of $S$ (that is, the number of
$|+\rangle $ states in eq.\ 7), and define
\begin{equation}
|k\rangle = \left( \frac{1}{N_k} \right)^{\frac{1}{2}} \sum _{n(S)=k} 
        |S\rangle .
\end{equation}
The number of (mutually-orthogonal) terms on the RHS of this equation is the
binomial coefficient $N!/[k!(N-k)!]$,
and so $|k\rangle $ is normalized if $N_{k} = N!/[k!(N-k)!]$. Note that
each term in the sum in eq.\  8, and hence $|k\rangle $ itself, is an
eigenstate of $M_x$, with eigenvalue $[k(+1) + (N-k)(-1)]/N$; 
\linebreak I will write this eigenvalue as $\lambda_{k}=(2k/N)-1$.

 From eqs.\ 1, 2, 7  and 8  , it is easy to see that
\begin{equation}
|\Psi \rangle = \sum_{k=0}^{N} c_{+}^{k}c_{-}^{(N-k)}N_{k}^{\frac{1}{2}}
        |k\rangle,
\end{equation} from which it follows that
\begin{equation}
|\langle k|\Psi \rangle |^{2} = |c_{+}|^{2k}|c_{-}|^{2(N-k)}
          \left(\frac{N!}{k!(N-k)!}\right).
\end{equation}
Since $|c_{-}|^{2} = 1 - |c_{+}|^{2},$ this shows that
$|\langle k|\Psi \rangle |^{2}$ is a binomial distribution in $k$.
For large values of $N$, the maximum of this distribution is \cite{F}
at $k=|c_{+}|^{2}N$ (and so $\lambda_{k}= \bar{\sigma}_x$), 
and Stirling's formula then shows that this
maximum value is $O(N^{-\frac{1}{2}})$.  This is Squires' result; 
roughly speaking, this result follows from the fact that, although 
$|\langle k|\Psi \rangle |^{2}$ is indeed peaked at $k=|c_{+}|^{2}N$,
the width of this peak grows as $N^{\frac{1}{2}}$.  Thus there are
$N^{\frac{1}{2}}$ terms which contribute significantly to the sum in
eq.\ 9, and so the overlap of $|\Psi \rangle$ with any one of them
must fall as $N^{-\frac{1}{4}}$. 

So it is not true that $|\Psi \rangle$ becomes close to an exact eigenstate
of $M_x$ as $N \rightarrow \infty $.  However, it could be said to 
come close to an approximate eigenstate, in the following sense:
suppose, for any $\epsilon >0$, we keep only those terms in eq.\ 9
in which $k$ is within $\epsilon N$ of the peak value (which is 
$N|c_{+}|^2$); that is,  let $a_{\pm} = N(|c_{+}|^2 \pm \epsilon )$,
and then define the state $|\Psi_{\epsilon}\rangle $ 
(up to normalization) by
\begin{equation}
|\Psi _{\epsilon} \rangle = \sum_{k=a_-}^{a_+} c_{+}^{k}c_{-}^{(N-k)}N_{k}^
        {\frac{1}{2}}|k\rangle.
\end{equation}
Since each term on the RHS of eq.\  11 is an eigenstate of $M_x$ 
with eigenvalue $\lambda_{k}$,
it follows from the non-probabilistic part of quantum mechanics \cite{Pos} 
that a measurement of $M_x$ on the state $|\Psi _{\epsilon} \rangle$
would certainly have an outcome which was in the range from 
$\lambda_{a_{-}}$ to $\lambda_{a_{+}}$, that is, from 
$\bar{\sigma}_x-2\epsilon$ to $\bar{\sigma}_x+2\epsilon$.
Also, since from eq.\ 10 it follows (see ref.\ \cite{F}) that
\begin{equation}
\sum_{k=a_-}^{a_+} |\langle k|\Psi \rangle |^{2} \rightarrow 1
\; \; \; \; \; \; \; {\rm as} \; N \rightarrow \infty ;
\end{equation}
then \cite{delt}
\begin{equation}
||\: |\Psi \rangle -|\Psi_{\epsilon} \rangle \: || \rightarrow 0
\; \; \; \; \; \; \; {\rm as} \; N \rightarrow \infty .
\end{equation}
So for any $\epsilon >0$, it is possible to define a state
$|\Psi_{\epsilon} \rangle$ on which a measurement of $M_x$
is certain to have  an outcome within $2\epsilon$ of  $\bar{\sigma}_x$
and which is close to $|\Psi\rangle$ for large $N$. 
The ``stability assumption'' of AR does not, as stated, specify precisely
how stable the results of experiments are expected to be, nor whether
that stability is uniform in $N$.  Nevertheless, if one is willing to
use that assumption, one could apply it directly to the state 
$|\Psi \rangle $ in the same spirit as AR apply it to the post-measurement
state.  Since a measurement of $M_x$ on the state
$|\Psi_{\epsilon} \rangle$ is certain to yield an outcome 
close to $\bar{\sigma}_x$,
and since  $|\Psi \rangle $ is close to 
$|\Psi_{\epsilon} \rangle$ for large $N$,
that assumption would seem to imply that the outcome of a measurement of 
$M_x$ on the state $|\Psi \rangle $ would almost certainly be close to
$\bar{\sigma} _x$.

Therefore, if one accepts the AR ``stability assumption'', 
it is not necessary to consider
the post-measurement state at all; the quantum probability rule
would follow from an application of this assumption to the 
pre-measurement state $|\Psi\rangle$.  Of course, this does not mean that
one could not obtain some insight from discussing, as 
AR have  done, the post-measurement state.

\vspace{1cm}
Acknowledgement: I would  like to acknowledge the hospitality of the
Lawrence Berkeley National Laboratory, where this work was done.


\begin{thebibliography}{9}
\bibitem{Pos} In fact, this follows from the postulate that a measurement on
an eigenstate of the measured operator will surely yield the 
corresponding eigenvalue.  Just define an operator $Q=\prod _{k=a{-}}
^{a_{+}}(M_{x}-\lambda _{k})$; the state $|\Psi _{\epsilon}\rangle$ 
is an eigenstate
of $Q$ with eigenvalue $0$, so a measurement of $Q$ on the state
$|\Psi _{\epsilon}\rangle$ would surely yield the value $0$.  But one way to
measure $Q$ is to measure $M_{x}$, so a measurement of $M_x$ must yield
one of the $\lambda _{k}$ with $k$ between $a_-$ and $a_+$.
\bibitem{delt} Relation (13) can also be seen to follow, without invoking
any properties of the binomial distribution, from the fact that the
norm of $|\Delta\rangle$ (defined in eq.\ 4) vanishes as 
$N\rightarrow \infty$. And since $|\langle k|\Psi \rangle |^2$
{\em is} a binomial distribution, then
$||\: |\Psi \rangle -|\Psi_{\epsilon} \rangle \: || \rightarrow 0$
faster than any inverse power of $N$.
\end{thebibliography}
\end{document}